\documentclass[10pt,english,a4paper,showpacs,amsmath,amssymb,floatfix,twocolumn]{revtex4-1}
\usepackage{amsmath,amstext,amssymb,graphicx,babel,geometry,setspace,bm,dcolumn,color}
\begin{document}

\title{Markovian evolution of classical and quantum correlations in transverse-field $XY$ model}
\author{Amit Kumar Pal}
\email{ak.pal@bosemain.boseinst.ac.in}
\author{Indrani Bose}
\email{indrani@bosemain.boseinst.ac.in}
\affiliation{Department of Physics, Bose Institute, 93/1, Acharya Prafulla Chandra Road, Kolkata - 700009, India}

\begin{abstract}
The transverse-field $XY$ model in one dimension is a well-known spin model for which the ground state 
properties and excitation spectrum are known exactly. The model has an interesting phase 
diagram describing quantum phase transitions (QPTs) belonging to two different universality classes. 
These are the transverse-field Ising model and the $XX$ model universality classes with both the models 
being special cases of the transverse-field $XY$ model. In recent years, quantities related to 
quantum information theoretic measures like entanglement, quantum discord (QD) and fidelity 
have been shown to provide signatures of QPTs. Another interesting issue is that of decoherence to 
which a quantum system is subjected due to its interaction, represented by a quantum channel, with 
an environment. In this paper, we determine the dynamics of different types of correlations present in a 
quantum system, namely, the mutual information $I\left(\rho_{AB}\right)$, the classical 
correlations $C\left(\rho_{AB}\right)$ and the quantum correlations $Q\left(\rho_{AB}\right)$, as 
measured by the quantum discord, in a two-qubit state. The density matrix of this state is given by  
the nearest-neighbour reduced density matrix obtained from the ground state of the transverse-field
$XY$ model in $1d$. We assume Markovian dynamics for the time-evolution due to system-environment 
interactions. The quantum channels considered include the bit-flip, bit-phase-flip and phase-flip
channels. Two different types of dynamics are identified for the channels in one of which the quantum 
correlations are greater in magnitude than the classical correlations in a finite time interval. 
The origins of the different types of dynamics are further explained. For the different channels, 
appropriate 
quantities associated with the dynamics of the correlations are identified which provide signatures 
of QPTs.  We also report results for further-neighbour two-qubit states and finite temperatures.
\end{abstract}

\pacs{75.10.Pq, 64.70.Tg, 03.67.-a, 03.65.Yz}

\maketitle

\section{Introduction}
\label{intro}

\begin{figure}
 \includegraphics[scale=0.65]{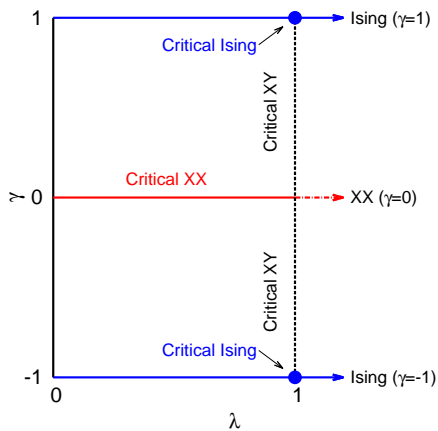}
 \caption{(Color online) Criticality of the transverse-field $XY$ model in the $(\lambda,\gamma)$ parameter space. 
         The transverse-field Ising model
         $(\gamma=\pm1)$ has a QCP at $\lambda_{c}=1$. The line $\lambda=1$ 
         represents criticality of the $XY$ model with universality class the same as that of 
         the transverse-field Ising model.  
         The $\gamma=0$ line ($XX$ model) represents the anisotropy transition line
         for $\lambda\in[0,1]$ with the transition belonging to a new universality class. }
 \label{phased}
\end{figure}

The fully anisotropic transverse-field $XY$ model in one dimension (1d) describes an interacting spin 
system for which many exact results on ground and excited state properties including spin correlations
are known \cite{mattis,mccoy,pfeuty}. The Hamiltonian describing the model is given by 
\begin{eqnarray}
H_{XY}&=&-\frac{\lambda}{2}\sum_{i=1}^{L}
\left\{(1+\gamma)\sigma_{i}^{x}\sigma_{i+1}^{x}+(1-\gamma)\sigma_{i}^{y}\sigma_{i+1}^{y}\right\}\nonumber\\
&&-\sum_{i=1}^{L}\sigma_{i}^{z}
\label{HXY}
\end{eqnarray}
where $L$ denotes the number of sites in the 1d lattice, $\sigma_{i}^{x,y,z}$ are the Pauli spin operators
defined at the lattice site $i$, $\gamma$ is the degree of anisotropy ($-1\leq\gamma\leq 1$) and 
$\lambda$ the inverse of the strength of the transverse magnetic field in the $z$ direction 
($\lambda>0$). The Hamiltonian is translation invariant and satisfies periodic boundary conditions. 
Two specific cases of the $XY$ model are the transverse field Ising model corresponding to $\gamma=\pm1$
and the isotropic $XX$ model ($\gamma=0$) in the transverse magnetic field. For all values of the anisotropy
parameter, the Hamiltonian can be diagonalized exactly in the thermodynamic limit $L\rightarrow \infty$
\cite{mattis,mccoy,pfeuty} through the successive applications of the Jordan-Wigner and Bogoliubov 
transformations. For 
non-zero values of $\gamma$, the model exhibits a second-order quantum phase transition (QPT) at the 
critical point $\lambda_{c}=1$ separating a ferromagnetic ordered phase from a quantum paramagnetic 
phase. The order parameter for the transition is $\langle\sigma^{x}\rangle$, the magnetization in the 
$x$ direction, which has a non-zero expectation value for $\lambda>1$. The magnetization in the 
$z$ direction, $\langle\sigma^{z}\rangle$, is non-zero for all values of $\lambda$ with its 
first derivative exhibiting a singularity at the critical point $\lambda_{c}=1$. In the interval 
$0<|\gamma|\leq1$, the critical point transition at $\lambda_{c}=1$ belongs to the Ising universality
class. For $\lambda\in[0,1]$, there is another QPT, termed the anisotropy transition, at the 
critical point $\gamma=0$. The transition belongs to a different universality class and separates two
ferromagnetic phases with orderings in the $x$ and $y$ directions respectively 
\cite{mattis,mccoy,pfeuty,zhong,dutta}. Figure \ref{phased} shows the phase diagram of the transverse-field $XY$
chain in the $(\lambda,\gamma)$ parameter space. 

The transverse-field $XY$ model and its special case, the transverse-field Ising model (TIM), are 
well-known statistical mechanical models which illustrate QPTs. A QPT occurs at $T=0$ and is brought 
about by tuning a non-thermal parameter, e.g., exchange interaction strength, external magnetic field
etc. In recent years, quantum information-related measures like entanglement 
\cite{osborne,osterloh,amico,lewenstein}, discord \cite{sarandy,werlang} 
and fidelity \cite{zanardi} have been used as indicators of QPTs. A first-order QPT is characterized 
by a discontinuity in the first derivative of the ground state energy with respect to the tuning 
parameter. Similarly in a second-order QPT, termed a critical point QPT, the second derivative of the 
ground state energy becomes discontinuous or divergent at the critical point. Entanglement and 
quantum discord (QD) provide different measures of quantum correlations in the mixed state of an 
interacting many-body system. The QD has so far been defined in the case of a bipartite quantum system. At a 
first-order QPT point, appropriate entanglement measures and the QD are known to become discontinuous
\cite{bose,wu} whereas a critical-point QPT is signaled by the discontinuity or divergence of the first 
derivative of the quantum correlation measures with respect to the tuning parameter \cite{wu}. The 
correlations between the different constituents of an interacting quantum system have two distinct 
components: classical and quantum. While entanglement provides the most well-known example of quantum 
correlations, the QD captures quantum correlations which are more general than those measured 
by entanglement \cite{ollivier,henderson}. In fact, there are separable mixed states which by definition are unentangled
but have non-zero QD.

Quantum systems, in general, are open systems because of the inevitable interaction of a system with its
environment. This results in decoherence accompanied by a decay of the quantum correlations in 
composite systems. A number of recent studies \cite{maziero,almeida,boas,vedral,mazzola,pal_clust,pal_ising} 
explore the dynamics of entanglement and QD
when the system-environment interactions are taken into account. An interesting result 
yielded by such studies is that the QD is more robust than entanglement in the case of Markovian
(memoryless) dynamics. The QD decays in time vanishing only asymptotically 
\cite{boas,mazzola,pal_clust,pal_ising,ferraro}
whereas entanglement undergoes a \textquoteleft sudden death\textquoteright \cite{maziero,almeida}, i.e.,
a complete disappearance at a finite time. The classical correlations, $C\left(\rho_{AB}\right)$, 
and the QD, $Q\left(\rho_{AB}\right)$, have mostly been computed only for two-qubit systems 
described by the density matrix $\rho_{AB}$ with $A$ and $B$ denoting the subsystems. Three general types
of Markovian time evolution have been observed \cite{vedral}: \textit{(i)} $C\left(\rho_{AB}\right)$ 
is constant as a function of time and $Q\left(\rho_{AB}\right)$ decays monotonically, \textit{(ii)}
$C\left(\rho_{AB}\right)$ decays monotonically over time till a parametrized time $p_{sc}$ is reached
and remains constant thereafter.  $Q\left(\rho_{AB}\right)$ has an abrupt change in the decay 
rate at $p_{sc}$, and has magnitude greater than that of $C\left(\rho_{AB}\right)$ in a parametrized time interval, 
and \textit{(iii)} both 
$C\left(\rho_{AB}\right)$ and $Q\left(\rho_{AB}\right)$ decay monotonically. Mazzola et al. \cite{mazzola}
have obtained a significant result that under Markovian dynamics and for a class of initial states 
the QD remains constant in a finite time interval $0<t<\tilde{t}$. In this time interval, the classical 
correlations $C\left(\rho_{AB}\right)$ decay monotonically. Beyond $t=\tilde{t}$, 
$C\left(\rho_{AB}\right)$ becomes constant whereas the QD decreases monotonically as a function of 
time. The sudden changes in the decay rates of correlations and their constancy in certain time 
intervals have been demonstrated in actual experiments \cite{xu,auccaise}.

In this paper, we consider a two-qubit system each qubit of which interact with an independent 
reservoir. The density matrix of the two-qubit system is described by the reduced density matrix 
derived from the ground state of the transverse-field $XY$ model in 1d. A similar study has 
earlier been carried out for the TIM in 1d \cite{pal_ising}. We investigate the dynamics of the 
QD as well as the classical correlations under Markovian time evolution and identify new features 
close to the quantum critical points of the model. The study is further extended to the cases of 
further-neighbour two-qubit states and finite temperatures. 
In Sec. \ref{scheme}, the calculational scheme for computing $C\left(\rho_{AB}\right)$ and 
$Q\left(\rho_{AB}\right)$ is described. We further discuss the methodology for studying the dynamics 
of the classical and quantum correlations for different quantum channels representing the 
system-environment interactions. Sec. \ref{results} presents the major results obtained and the discussions 
thereof. Sec. \ref{conclu} contains some concluding remarks.

\section{Classical and quantum correlations: Definition and dynamics}   
\label{scheme}

In classical information theory, the total correlation between two random variables $A$ and $B$ is 
quantified by their mutual information $I(A,B)=H(A)+H(B)-H(A,B)$. The random variables $A$ and $B$
take on the values \textquoteleft$a$\textquoteright$\;$ and \textquoteleft$b$\textquoteright$\;$ 
respectively with probabilities given by the sets $\left\{p_{a}\right\}$ and $\left\{p_{b}\right\}$.
$H(A)=-\sum_{a}p_{a}\log_{2}p_{a}$ and $H(B)=-\sum_{b}p_{b}\log_{2}p_{b}$ are the Shannon entropies 
for the variables $A$ and $B$. $H(A,B)=-\sum_{a,b}p_{a,b}\log_{2}p_{a,b}$ is the joint Shannon 
entropy for the variables $A$ and $B$ with $p_{a,b}$ being the joint probability that the variables 
$A$ and $B$ have the respective values $a$ and $b$. Generalization to the quantum case is straightforward 
with the density matrix $\rho$ replacing the classical probability distribution and 
the von Neumann entropy $S(\rho)=-\mbox{Tr}\left(\rho\log_{2}\rho\right)$ replacing the Shannon 
entropy. The expression for the quantum mutual information is given by 
\begin{equation}
 I\left(\rho_{AB}\right)=S\left(\rho_{A}\right)+S\left(\rho_{B}\right)-S\left(\rho_{AB}\right)
\label{mutual1}
\end{equation}
An equivalent expression for the classical mutual information is $J(A,B)=H(A)-H(A|B)$ where the 
conditional entropy $H(A|B)$ is a measure of our ignorance about the state of $A$ when that of $B$
is known. The equivalence with the earlier expression for the classical mutual information can 
be shown via the Bayes' rule. The quantum versions of $J(A,B)$ and $I(A,B)$ are not, however, equivalent 
because the magnitude of the quantum conditional entropy is determined by the type of measurement
performed on the subsystem $B$ to gain knowledge of its state. Different measurement choices yield 
different values for the conditional entropy. We consider von Neumann-type measurements on $B$ 
represented by a complete set of orthogonal projectors, $\left\{\Pi_{k}^{B}\right\}$, corresponding 
to the set of possible outcomes $k$. The final state of the composite quantum system after 
measurement on $B$ is given by  
$\rho_{k}=\left(I\otimes\Pi_{k}^{B}\right)\rho_{AB}\left(I\otimes\Pi_{k}^{B}\right)/p_{k}$
with $p_{k}=\mbox{Tr} \left(\left(I\otimes\Pi_{k}^{B}\right)
\rho_{AB}\left(I\otimes\Pi_{k}^{B}\right)\right)$. $I$ denotes the identity operator for the 
subsystem $A$ and $p_{k}$ is the probability for obtaining the outcome $k$. The quantum conditional 
entropy is given by   
\begin{equation}
 S\left(\rho_{AB}\left|\left\{ \Pi_{k}^{B}\right\} \right.\right)=\sum_{k}p_{k}S\left(\rho_{k}\right)
\label{qce}
\end{equation} 
leading to the following quantum extension of the classical mutual information
\begin{equation}
J\left(\rho_{AB},\left\{ \Pi_{k}^{B}\right\} \right)
=S\left(\rho_{A}\right)-S\left(\rho_{AB}\left|\left\{ \Pi_{k}^{B}\right\} \right.\right)
\label{mutual2}
\end{equation}
The projective measurements on the subsystem $B$ removes all the non-classical correlations between 
the subsystems $A$ and $B$. Henderson and Vedral \cite{henderson} have quantified the total 
classical correlations $C\left(\rho_{AB}\right)$ by maximizing 
$J\left(\rho_{AB},\left\{ \Pi_{k}^{B}\right\} \right)$ w.r.t. $\left\{\Pi_{k}^{B}\right\}$, i.e., 
\begin{equation}
C\left(\rho_{AB}\right)
=\underset{\left\{ \Pi_{k}^{B}\right\} }{\max}
\left(J\left(\rho_{AB},\left\{ \Pi_{k}^{B}\right\} \right)\right)
\label{classical}
\end{equation}
The difference between the total correlations $I\left(\rho_{AB}\right)$
(Eq. (\ref{mutual1})) and the classical correlations $C\left(\rho_{AB}\right)$ (Eq. (\ref{classical})) 
defines the QD, 
$Q\left(\rho_{AB}\right)$, 
\begin{equation}
Q\left(\rho_{AB}\right)=I\left(\rho_{AB}\right)-C\left(\rho_{AB}\right)
\label{discord}
\end{equation}

\begin{figure}
 \includegraphics[scale=0.65]{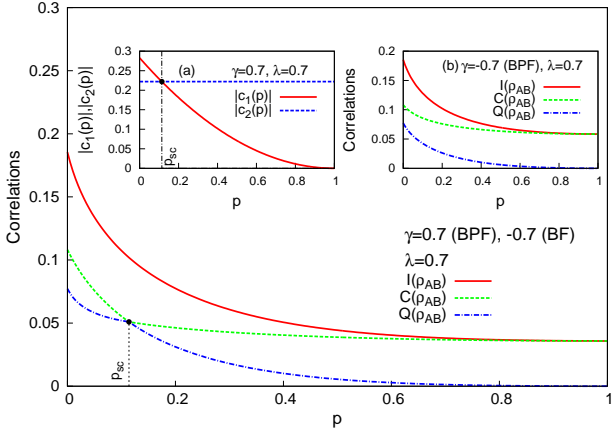}
 \caption{(Color online) Bit-phase-flip channel: Decay of mutual information 
         $I\left(\rho_{AB}\right)$ (solid line), classical correlations $C\left(\rho_{AB}\right)$
         (dashed line) and the 
         QD $Q\left(\rho_{AB}\right)$ (dot-dashed line)
         as a function of the parametrized time 
         $p=1-e^{-\theta t}$ for $\lambda=0.7,\gamma=0.7$. Also, $p_{sc}=0.114$. (inset)
         (i)The variations of $\left|c_{1}(p)\right|$ (solid line) and 
         $\left|c_{2}(p)\right|$ (dashed line) as functions 
         of $p$. The crossing of $\left|c_{1}(p)\right|$ with $\left|c_{2}(p)\right|$ sets the 
         parametrized time point $p_{sc}$. 
         (ii)Decay of mutual information 
         $I\left(\rho_{AB}\right)$ (solid line), classical correlations $C\left(\rho_{AB}\right)$ 
         (dashed line) and 
         quantum discord $Q\left(\rho_{AB}\right)$ (dot-dashed line)  
         as a function of the parametrized time 
         $p=1-e^{-\theta t}$ for $\lambda=0.7,\gamma=-0.7$.
         }
 \label{bpfp}
\end{figure}

Though the concept of the QD is well-established, its computation is restricted to mostly 
two-qubit states. For the two-qubit $X$ states, analytic expressions for the QD can be derived 
when the density matrix, defined in the computational basis  
$\left\{|11\rangle,|10\rangle,|01\rangle,|00\rangle\right\}$, has the following structure
\begin{eqnarray}
\rho_{AB}= \left(
 \begin{array}{cccc}
 a & 0 & 0 & f \\
 0 & b & z & 0 \\
 0 & z & b & 0 \\
 f & 0 & 0 & d 
 \end{array}
 \right)
 \label{dm}
\end{eqnarray}
where $A$ and $B$ correspond to the two individual qubits and $z$ and $f$ are real numbers. The 
eigenvalues of $\rho_{AB}$ are \cite{sarandy}
\begin{eqnarray}
 \lambda_{0}&=&\frac{1}{4}\left\{\left(1+c_{3}\right)+\sqrt{4 c_{4}^{2}+\left(c_{1}-c_{2}\right)^{2}}\right\} \nonumber \\
 \lambda_{1}&=&\frac{1}{4}\left\{\left(1+c_{3}\right)-\sqrt{4 c_{4}^{2}+\left(c_{1}-c_{2}\right)^{2}}\right\} \nonumber \\
 \lambda_{2}&=&\frac{1}{4}\left(1-c_{3}+c_{1}+c_{2}\right) \nonumber \\
 \lambda_{3}&=&\frac{1}{4}\left(1-c_{3}-c_{1}-c_{2}\right)
 \label{ev}
\end{eqnarray}
with 
\begin{eqnarray}
 c_{1}&=&2z+2f \nonumber \\
 c_{2}&=&2z-2f \nonumber \\
 c_{3}&=&a+d-2b \nonumber \\
 c_{4}&=&a-d 
 \label{cv}
\end{eqnarray}
The mutual information $I\left(\rho_{AB}\right)$ can be written as \cite{sarandy}
\begin{eqnarray}
 I\left(\rho_{AB}\right)=S\left(\rho_{A}\right)+S\left(\rho_{B}\right)
                         +\sum_{\alpha=0}^{3}\lambda_{\alpha}\log_{2}\lambda_{\alpha}
\label{mi}
\end{eqnarray}
where 
\begin{eqnarray}
 S\left(\rho_{A}\right)=S\left(\rho_{B}\right)&=&-\frac{1+c_{4}}{2}\log_{2}\frac{1+c_{4}}{2}\nonumber \\
&&-\frac{1-c_{4}}{2}\log_{2}\frac{1-c_{4}}{2}
\label{vne}                                               
\end{eqnarray}
With expressions for $I\left(\rho_{AB}\right)$ and $C\left(\rho_{AB}\right)$ given in Eqs. 
(\ref{mi}), (\ref{vne}) and (\ref{classical}), the QD, $Q\left(\rho_{AB}\right)$, (Eq. (\ref{discord}))
can in principle be computed. The difficulty lies in carrying out the maximization procedure needed
for the computation of $C\left(\rho_{AB}\right)$. It is possible to do so analytically when 
$\rho_{AB}$ is of the form given in (\ref{dm}) resulting in the following expressions for the 
QD \cite{fanchini}: 

\begin{eqnarray}
 Q\left(\rho_{AB}\right)=\mbox{min}\left\{Q_{1},Q_{2}\right\}
\label{qd}
\end{eqnarray}
where 
\begin{eqnarray}
 Q_{1}&=&S\left(\rho_{B}\right)-S\left(\rho_{AB}\right)-a\log_{2}\frac{a}{a+b}\nonumber \\
&&-b\log_{2}\frac{b}{a+b}
-d\log_{2}\frac{d}{d+b}-b\log_{2}\frac{b}{d+b}
 \label{qd1}
\end{eqnarray}
and
\begin{eqnarray}
 Q_{2}&=&S\left(\rho_{B}\right)-S\left(\rho_{AB}\right)-\Delta_{+}\log_{2}\Delta_{+}\nonumber \\
 &&-\Delta_{-}\log_{2}\Delta_{-}
 \label{qd2}
\end{eqnarray}
with $\Delta_{\pm}=\frac{1}{2}\left(1\pm\Gamma \right)$ and 
$\Gamma^{2}=\left(a-d\right)^{2}+4\left(|z|+|f|\right)^{2}$

For spin Hamiltonians with certain symmetries, the two-spin reduced density matrix $\rho_{ij}$
has a form similar to that given in Eq. (\ref{dm}) (the qubits $A$ and $B$ now 
represent the spins at sites $i$ and $j$ respectively) where the matrix elements of 
$\rho_{ij}$ can be written in 
terms of single-site magnetization and two-site spin correlation functions. In the case of the 
$XY$ model in a transverse-field, the matrix elements of the two-site reduced density matrix 
are \cite{dutta,osborne,dillenschneider}
\begin{eqnarray}
 a&=&\frac{1}{4}+\frac{\langle\sigma^{z}\rangle}{2}+
     \frac{\langle\sigma_{i}^{z}\sigma_{j}^{z}\rangle}{4} \nonumber \\
 d&=&\frac{1}{4}-\frac{\langle\sigma^{z}\rangle}{2}+
     \frac{\langle\sigma_{i}^{z}\sigma_{j}^{z}\rangle}{4} \nonumber \\
 b&=&\frac{1}{4}\left(1-\langle\sigma_{i}^{z}\sigma_{j}^{z}\rangle\right) \nonumber \\
 z&=&\frac{1}{4}
     \left(\langle\sigma_{i}^{x}\sigma_{j}^{x}\rangle+\langle\sigma_{i}^{y}\sigma_{j}^{y}\rangle\right) \nonumber \\
 f&=&\frac{1}{4}
     \left(\langle\sigma_{i}^{x}\sigma_{j}^{x}\rangle-\langle\sigma_{i}^{y}\sigma_{j}^{y}\rangle\right) 
 \label{elements}
\end{eqnarray}
The finite temperature single-site magnetization $\langle\sigma^{z}\rangle$ in the case of the transverse-field 
$XY$ model is given by \cite{dutta,dillenschneider} 
\begin{eqnarray}
 \langle\sigma^{z}\rangle=-\frac{1}{\pi}\int_{0}^{\pi}d\phi\tanh{\left(\frac{\beta \omega_{\phi}}{2}\right)}\frac{\left(1+\lambda\cos{\phi}\right)}
                           {\omega_{\phi}}
\end{eqnarray}
where 
\begin{eqnarray}
 \omega_{\phi}=\sqrt{\left(\gamma\lambda\sin{\phi}\right)^{2}+\left(1+\lambda\cos{\phi}\right)^{2}}
\end{eqnarray}
describes the energy spectrum and $\beta=\frac{1}{kT}$ with $k$ being the Boltzman constant and $T$ the 
absolute temperature. The spin-spin correlation functions are obtained from the determinants of 
Toeplitz matrices \cite{dillenschneider,mccoy,pfeuty} as
\begin{eqnarray}
  \left\langle\sigma_{i}^{x}\sigma_{i+r}^{x}\right\rangle &=& 
 \left|
 \begin{array}{cccc}
  G_{-1} & G_{-2} & \cdots & G_{-r}\\
  G_{0} & G_{-1} & \cdots & G_{-r+1}\\
  \vdots & \vdots & \ddots & \vdots\\
  G_{r-2} & G_{r-3} & \cdots & G_{-1}\\
 \end{array}\right| \nonumber \\
  \left\langle\sigma_{i}^{y}\sigma_{i+r}^{y}\right\rangle &=& 
 \left|
 \begin{array}{cccc}
  G_{1} & G_{0} & \cdots & G_{-r+2}\\
  G_{2} & G_{1} & \cdots & G_{-r+3}\\
  \vdots & \vdots & \ddots & \vdots\\
  G_{r} & G_{r-1} & \cdots & G_{1}\\
 \end{array}\right| \nonumber \\
 \left\langle\sigma_{i}^{z}\sigma_{i+r}^{z}\right\rangle &=&
 \left\langle\sigma^{z}\right\rangle^{2}-G_{r}G_{-r}
\label{toepliz} 
\end{eqnarray}
where 
\begin{eqnarray}
  G_{r}&=&\frac{1}{\pi}\int_{0}^{\pi}d\phi\tanh{\left(\frac{\beta \omega_{\phi}}{2}\right)}\cos(r\phi)
 \frac{\left(1+\lambda\cos\phi\right)}{\omega_{\phi}}\nonumber \\
&&-\frac{\gamma\lambda}{\pi}\int_{0}^{\pi}d\phi\tanh{\left(\frac{\beta \omega_{\phi}}{2}\right)}\sin(r\phi)\frac{\sin\phi}{\omega_{\phi}}
 \label{spincor}
\end{eqnarray}
with $r=|i-j|$ being the distance between the two spins at the sites $i$ and $j$ (for the 
nearest-neighbour (n.n.) case, $r=1$).

\begin{figure}
 \includegraphics[scale=0.65]{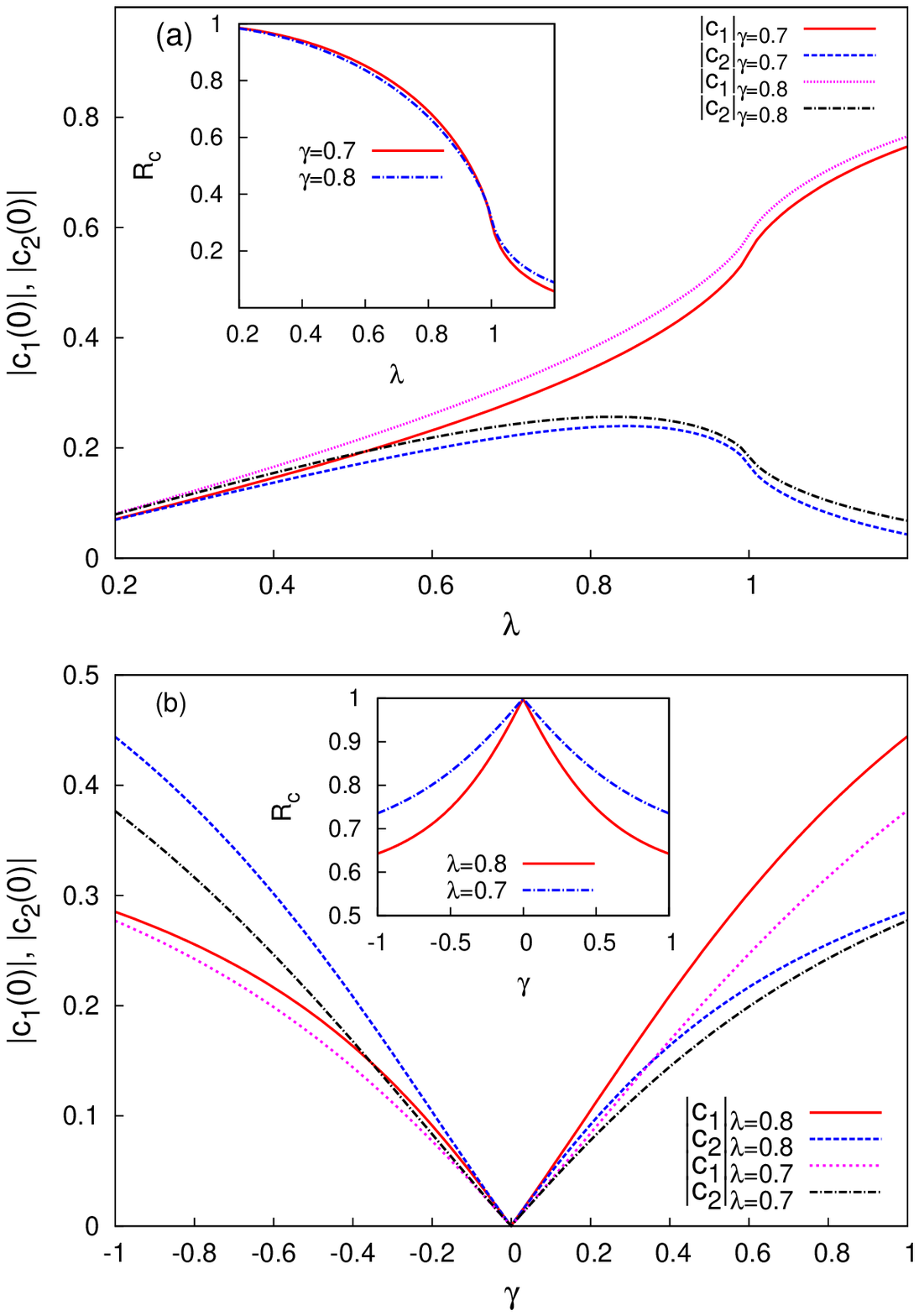}
 \caption{(Color online) (a) Variations of $\left|c_{1}(0)\right|$ and 
         $\left|c_{2}(0)\right|$ 
         with $\lambda$ for $\gamma=0.7$ ($\left| c_{1}(0)\right|$ solid line and $\left| c_{2}(0)\right|$ dashed line)
         and $\gamma=0.8$ ($\left| c_{1}(0)\right|$ dotted line and $\left| c_{2}(0)\right|$ dot-dashed line). 
         (inset) Variation of 
         $R_{c}$ $\left(=\left|c_{2}(0)\right|/\left|c_{1}(0)\right|\right)$ with $\lambda$ for 
         $\gamma=0.7$ (solid line) and $\gamma=0.8$ (dot-dashed line).
         (b)Variations of $\left|c_{1}(0)\right|$ and $\left|c_{2}(0)\right|$
         with $\gamma$ for $\lambda=0.7$ ($\left| c_{1}(0)\right|$ dotted line and $\left| c_{2}(0)\right|$ dot-dashed line) 
         and 
         $\lambda=0.8$ ($\left| c_{1}(0)\right|$ solid line and $\left|c_{2}(0)\right|$ dashed line). (inset) Variation of 
         $R_{c}$ with $\gamma$ for $\lambda=0.7$ (dot-dashed line) and $\lambda=0.8$ (solid line). 
         $R_{c}=\left|c_{2}(0)\right|/\left|c_{1}(0)\right|$ for $\gamma>0$ (BPF channel) and 
         $R_{c}=\left|c_{1}(0)\right|/\left|c_{2}(0)\right|$ for $\gamma<0$ (BF channel).}
 \label{cgl}
\end{figure}

We next consider the interaction of the chain of qubits, each qubit representing a $S=\frac{1}{2}$ spin
in transverse-field $XY$ model, with an environment. We choose the initial state (time $t=0$) of the whole 
system at time $t=0$ to be of the product 
form, i.e., 
\begin{eqnarray}
 \rho(0)=\rho_{s}(0)\otimes\rho_{e}(0)
 \label{sysden}
\end{eqnarray}
where the density matrices $\rho_{s}$ and $\rho_{e}$ correspond to the system and environment 
respectively. We assume that the environment is represented by $L$ independent reservoirs each of 
which interacts locally with a qubit constituting the $XY$ chain. The reduced density matrix 
obtained from Eq. (\ref{sysden}) can be written as
\begin{eqnarray}
 \rho_{r}(0)=\rho_{rs}(0)\otimes\rho_{re}(0)
\end{eqnarray}
where $\rho_{rs}$ and $\rho_{re}$ represent the two-qubit reduced density matrix of the 
transverse-field $XY$ chain and the corresponding reduced density matrix of the 
two-reservoir environment respectively. The quantum channel describing the interaction 
between a qubit and its environment can be of various types \cite{maziero,nielsenchuang}. 
We first assume the two qubits to represent nearest-neighbour spins in the $XY$ chain.
In this paper, we investigate the dynamics of the two-qubit classical
and quantum correlations (in the form of the QD) under the influence of the bit-flip (BF), 
bit-phase-flip (BPF) and phase-flip (PF) channels.
In the Kraus operator representation, an initial state, $\rho_{rs}(0)$, of the qubits evolves as 
\cite{boas,nielsenchuang}
\begin{eqnarray}
 \rho_{rs}(t)=\sum_{\mu,\nu}E_{\mu,\nu}\rho_{rs}(0)E_{\mu,\nu}^{\dagger}
\label{evolve}
\end{eqnarray}
where the Kraus operators $E_{\mu,\nu}=E_{\mu}\otimes E_{\nu}$ satisfy the completeness relation
$\sum_{\mu,\nu}E_{\mu,\nu}E_{\mu,\nu}^{\dagger}=I$ for all $t$.
The BF, BPF and PF channels destroy the information contained in the phase relations without 
involving an exchange of energy. The Kraus operators for these channels are
\begin{eqnarray}
 E_{0}=\sqrt{q^{\prime}}\left(
 \begin{array}{cc}
  1 & 0 \\
  0 & 1
 \end{array}\right);\;\;
 E_{1}=\sqrt{p/2}\sigma_{i}
\label{kraus}
\end{eqnarray}  
where $i=x$ for the BF, $i=y$ for the BPF and $i=z$ for the PF channel with
$q^{\prime}=1-p/2$.  In the case of Markovian time evolution, $p$ is given by  
$p=1-e^{-\theta t}$ with $\theta$ denoting the decay rate. 
A fuller description of the channels and the corresponding dynamics can be obtained 
from Refs. \cite{boas,pal_ising,nielsenchuang}. Using the two-spin reduced density matrix 
described by Eqs. (\ref{dm}) and (\ref{elements})-(\ref{spincor}) as the initial state 
$\rho_{rs}(0)$,  one can calculate the time-evolved state 
$\rho_{rs}(t)$ for a specific quantum channel using the Eqs. (\ref{evolve}) and (\ref{kraus}). 
Since $\rho_{rs}(t)$ has the same form as Eq. (\ref{dm}), 
the mutual information, $I(\rho_{AB})$, the QD $Q(\rho_{AB})$ and the 
classical correlations $C(\rho_{AB})$ can be computed at any time $t$ with the help of the formulae 
in Eqs. (\ref{ev})-(\ref{qd2}). The main results of our calculations for the various 
quantum channels are described in the following section.

\section{Results}
\label{results}

\subsection{Nearest-neighbour quantum correlations at zero temperature}

\subsubsection{Bit-flip and bit-phase-flip channels}
\label{bpfss}

\begin{figure}
 \includegraphics[scale=0.65]{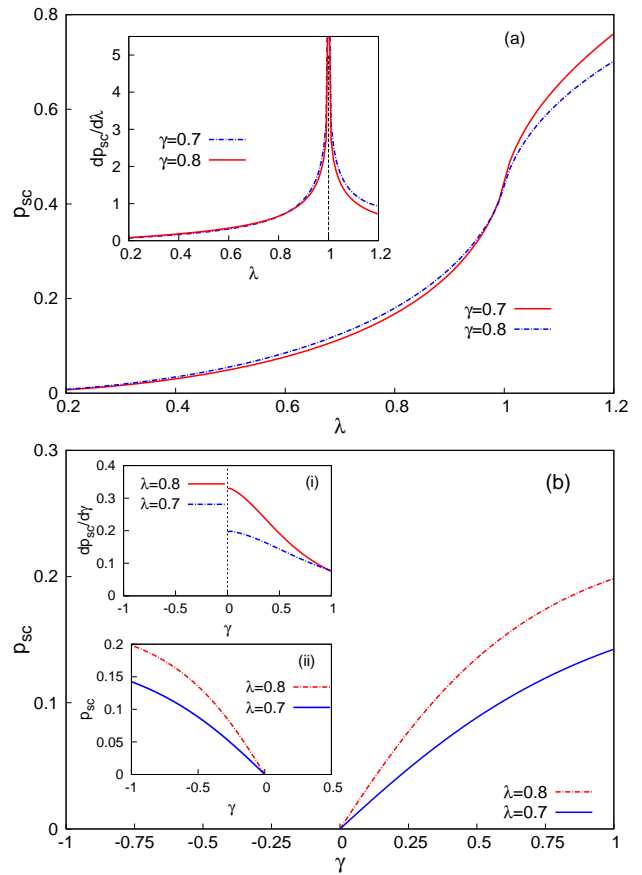}
 \caption{(Color online)(a) Variation of $p_{sc}$ as a function of $\lambda$ with $\gamma=0.7$ (solid line) and 
         $\gamma=0.8$ (dot-dashed line). (inset) The first derivative of $p_{sc}$ w.r.t. $\lambda$ diverges as 
         the QCP $\lambda_{c}=1$ is approached.
         (b)  Variation of $p_{sc}$ as a function of $\gamma$ with $\lambda=0.7$ (solid line) and 
         $\lambda=0.8$ (dot-dashed line) for the BPF channel. (inset) (i) The first derivative of $p_{sc}$ 
         w.r.t. $\gamma$ exists when $\gamma_{c}>0$ for the BPF channel. 
         (ii) Variation of $p_{sc}$ as functions of $\gamma$ with $\lambda=0.7$ (solid line) and 
         $\lambda=0.8$ (dot-dashed line) for the BF channel.}
 \label{pgl1}
\end{figure}

We first consider the case of $T=0$ so that the $XY$ spin chain is in its ground state. 
In the case of the BPF channel, 
the dynamical evolutions of the mutual information $I\left(\rho_{AB}\right)$, the classical 
correlations $C\left(\rho_{AB}\right)$, the QD $Q\left(\rho_{AB}\right)$ and the absolute
values of the coefficients $c_{1}$ and $c_{2}$ as a function of the 
parametrized time $p$ $(p=1-e^{-\theta t})$ are shown in Fig. \ref{bpfp} 
for $\lambda=\gamma=0.7$. The dynamics are similar to type \textit{(ii)} where sudden changes in 
the decay rates of the classical correlations and the QD occur at a 
parametrized time instant $p_{sc}$. 
Inset (a) shows the variations of $\left|c_{1}(p)\right|$ and $\left|c_{2}(p)\right|$ versus $p$.
The crossing point is given by $p_{sc}$. 
Inset (b) shows the time-evolutions of the mutual information, the
classical correlations and the QD  for $\lambda=0.7$, $\gamma=-0.7$ which are described by
type \textit{(iii)} dynamics, i.e., both $C\left(\rho_{AB}\right)$ and $Q\left(\rho_{AB}\right)$ 
decay monotonically. The dynamics in the case of the BF channel have an interesting 
correspondence with the dynamics of the BPF channel. Type \textit{(ii)} and Type \textit{(iii)} 
dynamics are obtained in the case of the BPF (BF) channel for +ve (-ve) and -ve (+ve) values 
of $\gamma$ respectively. The following analysis provides us with an understanding of this result. 
The rules of evolution for the coefficients $c_{1}$ and $c_{2}$ are obtained from Eqs. (\ref{dm}), 
(\ref{cv}), (\ref{evolve}) and (\ref{kraus}) in the cases of the BPF and BF channels. These are 
given by 

\noindent\textit{BPF Channel:}
\begin{eqnarray}
 c_{1}(p)&=&c_{1}(0)(1-p)^{2} \nonumber\\
 c_{2}(p)&=&c_{2}(0)\nonumber\\
 c_{3}(p)&=&c_{3}(0)(1-p)^{2} \nonumber\\
 c_{4}(p)&=&c_{4}(0)(1-p)
\label{bpft}
\end{eqnarray}

\noindent\textit{BF Channel:}
\begin{eqnarray}
 c_{1}(p)&=&c_{1}(0)\nonumber\\
 c_{2}(p)&=&c_{2}(0)(1-p)^{2}\nonumber\\
 c_{3}(p)&=&c_{3}(0)(1-p)^{2} \nonumber\\
 c_{4}(p)&=&c_{4}(0)(1-p)
\label{bft}
\end{eqnarray}
\noindent From Eqs. (\ref{cv}) and (\ref{elements}), the coefficients $c_{1}(0)$ and $c_{2}(0)$ 
are identified as the two-spin correlation functions $\langle\sigma_{i}^{x}\sigma_{j}^{x}\rangle$ and 
$\langle\sigma_{i}^{y}\sigma_{j}^{y}\rangle$ respectively (in the present study, $r=|i-j|=1$). 
Fig. \ref{cgl}(a) shows the variation of the absolute values of the coefficients $c_{1}(0)$ 
and $c_{2}(0)$ with $\lambda$ for $\gamma>0$. The general observation is that for $\gamma>0$, 
$\left|c_{1}(0)\right|\geq\left|c_{2}(0)\right|$ for $\lambda\in[0,1]$. Fig. \ref{cgl}(b) shows the 
plots of $\left|c_{1}(0)\right|$ and $\left|c_{2}(0)\right|$ versus $\gamma$ for fixed $\lambda$
values. One finds that when $\gamma$ is $<0$, $\left|c_{1}(0)\right|\leq\left|c_{2}(0)\right|$
for $\lambda\in[0,1]$. Since $\left|c_{1}(0)\right|\geq\left|c_{2}(0)\right|$ for $\gamma>0$, 
Eq. (\ref{bpft}), describing the rules of evolution for the BPF channel, indicates that in this case
there is a certain parametrized time instant at which $\left|c_{1}(p)\right|$ crosses 
$\left|c_{2}(p)\right|\left(=\left|c_{2}(0)\right|\right)$. This specific value turns out to be 
the same as $p_{sc}$ at which a sudden change in the decay rates of the classical and quantum 
correlations occur (Fig. \ref{bpfp}). From the condition 
$\left|c_{1}(p)\right|=\left|c_{2}(p)\right|$ and Eq. (\ref{bpft}), one obtains the following 
expression for $p_{sc}$ in the case of the BPF channel:
\begin{eqnarray}
  p_{sc}=1-\sqrt{R_{c}}
 \label{psc}
\end{eqnarray}
where $R_{c}=\left|c_{2}(0)\right|/\left|c_{1}(0)\right|$. The QD (BPF channel, $\gamma>0$) 
is given by $Q_{2}$ in Eq. (\ref{qd2}) for $p\in[0,1]$. 
At $p=p_{sc}$, 
$\left(c_{1}(p)-c_{2}(p)\right)$ or $\left(c_{1}(p)+c_{2}(p)\right)$ changes sign depending on 
whether $c_{1}(0)$ and $c_{2}(0)$ have the same or different signs respectively. 
Noticing that $z(p)=\left(c_{1}(p)+c_{2}(p)\right)/4$ and  $f(p)=\left(c_{1}(p)-c_{2}(p)\right)/4$
(Eq. (\ref{cv})), the magnitude of $\Gamma$ which involves $|z|$ and $|f|$ (Eq. (\ref{qd2}))
changes at the point $p=p_{sc}$, while $S\left(\rho_{B}\right)$ and $S\left(\rho_{AB}\right)$ 
remain unchanged, thus  bringing about the sudden change in the decay rates of the 
quantum correlations.    
With $\gamma<0$, $\left|c_{1}(0)\right|$ is $\leq\left|c_{2}(0)\right|$ so that $c_{1}(p)$ 
never crosses $c_{2}(p)$ and one obtains type \textit{(iii)} dynamics. We next compare the rules 
of evolution for the BF channel (Eq. (\ref{bft})) with those of the BPF channel (Eq. (\ref{bpft})).
One finds that the rules of evolution for $c_{3}(p)$ and $c_{4}(p)$ remain identical while those 
for $c_{1}(p)$ and $c_{2}(p)$ are interchanged. This explains the interchange of dynamics type 
mentioned earlier: for $\gamma>0$, since $\left|c_{1}(0)\right|$ is $\geq\left|c_{2}(0)\right|$, 
$c_{1}(p)$ never crosses $c_{2}(p)$ for the BF channel so that type \textit{(iii)} dynamics hold true whereas for 
$\gamma<0$, the crossing point is given by 
$R_{c}=\left|c_{1}(0)\right|/\left|c_{2}(0)\right|$. The effective interchange in the identities 
of $c_{1}(0)$ and $c_{2}(0)$ across the $\gamma=0$ (critical $XX$) line is understood by 
remembering that $c_{1}(0)=\langle\sigma_{i}^{x}\sigma_{i+1}^{x}\rangle$ and 
$c_{2}(0)=\langle\sigma_{i}^{y}\sigma_{i+1}^{y}\rangle$. The correlation function
$\langle\sigma_{i}^{x}\sigma_{i+1}^{x}\rangle$ is interchanged with
$\langle\sigma_{i}^{y}\sigma_{i+1}^{y}\rangle$ under the transformation $\gamma\rightarrow-\gamma$
(see Eqs. (\ref{toepliz}) and (\ref{spincor})). The variations of $R_{c}$ with $\lambda$ and 
$\gamma$ are shown in the insets of Fig. \ref{cgl}.

\begin{figure}
 \includegraphics[scale=0.65]{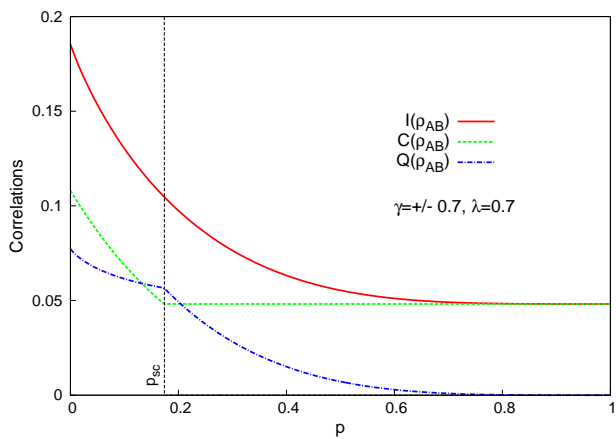}
 \caption{(Color online) Phase-flip channel: Decay of mutual information 
         $I\left(\rho_{AB}\right)$ (solid line), classical correlations $C\left(\rho_{AB}\right)$ 
         (dashed line) and 
         quantum discord $Q\left(\rho_{AB}\right)$ (dot-dashed line)  
         as a function of the parametrized time 
         $p=1-e^{-\theta t}$ for $\lambda=0.7,\gamma=\pm0.7$. In this case, $p_{sc}=0.173$.}
 \label{pfp}
\end{figure}

Fig. \ref{pgl1}(a) shows the plots of $p_{sc}$ and the first derivative of $p_{sc}$ w.r.t. $\lambda$, 
$\left(\frac{d p_{sc}}{d\lambda}\right)_{\gamma}$ (inset), as functions of $\lambda$
for the BPF channel $(\gamma>0)$.
The first derivative of $p_{sc}$ w.r.t. $\lambda$ diverges as the quantum critical point (QCP) $\lambda_{c}=1$ is approached.
Fig. \ref{pgl1}(b) shows the variation of $p_{sc}$ with the anisotropy parameter $\gamma$ 
for the BPF channel. As expected, $p_{sc}\neq0$ for $\gamma>0$. Both $\left|c_{1}\right|$ and $\left|c_{2}\right|$
individually tend to zero as $\gamma\rightarrow0$ for $\lambda\in[0,1]$ and the ratio 
$R_{c}\left(=\left|c_{2}(0)\right|/\left|c_{1}(0)\right|\right)\rightarrow1$ as $\gamma\rightarrow0$ 
(Fig. \ref{cgl}). The inset (i) shows the variation of the first derivative of $p_{sc}$ w.r.t. $\gamma$, 
$\left(\frac{d p_{sc}}{d\gamma}\right)_{\lambda}$, as a function of $\gamma$ for 
fixed values of $\lambda$. The first derivative of $p_{sc}$ w.r.t. $\gamma$ shows a discontinuity
at the anisotropy transition point, $\gamma=0$, indicating a QPT. Inset (ii) shows the 
variation of $p_{sc}$ as a function of $\gamma$ for the BF channel ($p_{sc}\neq0$ for $\gamma<0$). 
In a recent study \cite{pal_ising}, we have investigated the Markovian evolution of the 
classical and quantum correlations for a number of quantum channels in the case of the  transverse-field 
Ising model, a special case ($\gamma=1$) of the transverse-field $XY$ model considered in this paper.
For the BPF channel, we have pointed out that the first derivative of the  parametrized time instant 
at which the sudden changes in the decay rates of classical and quantum correlations occur w.r.t. 
$\lambda$ diverges as the Ising model critical point $\lambda_{c}=1$ is approached. We have further 
shown that the BPF (BF) channel exhibits only type \textit{(ii)} (type \textit{(iii)}) dynamics.
In the case of the $XY$ model, positive and negative values of the 
anisotropy parameter $\gamma$ allow for both type \textit{(ii)} ($\gamma<0$) and \textit{(iii)}
($\gamma>0$) dynamics for the BF channel. 
Similarly, unlike in the case of the transverse-field Ising model, 
type \textit{(iii)} dynamics are possible for the BPF channel ($\gamma<0$).

\subsubsection{Phase-flip channel}
\label{pfss}

\begin{figure}
 \includegraphics[scale=0.65]{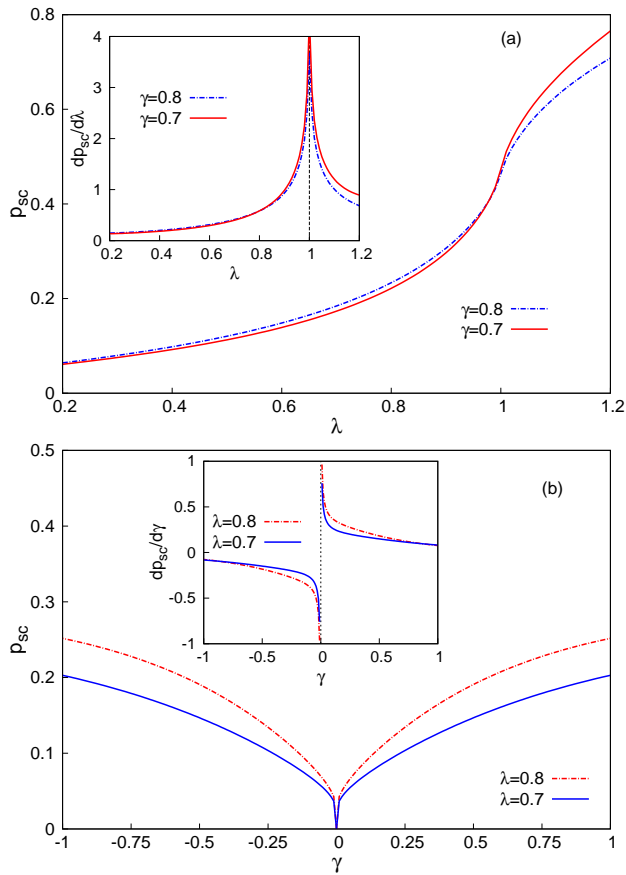}
 \caption{(Color online)(a) Variation of $p_{sc}$ as functions of $\lambda$ with $\gamma=0.7$ 
         (solid line) and 
         $\gamma=0.8$ (dot-dashed line) in the case of the PF channel. (inset) The first derivative of 
         $p_{sc}$ w.r.t. $\lambda$ diverges as 
         the QCP $\lambda_{c}=1$ is approached.
         (b)  Variation of $p_{sc}$ as functions of $\gamma$ with $\lambda=0.7$ (solid line) and 
         $\lambda=0.8$ (dashed line) for the PF channel. (inset) The first derivative of $p_{sc}$ 
         as a function of $\gamma$ becomes discontinuous at the anisotropy transition point $\gamma_{c}=0$.}
 \label{pgl2}
\end{figure}

\begin{figure}
 \includegraphics[scale=0.65]{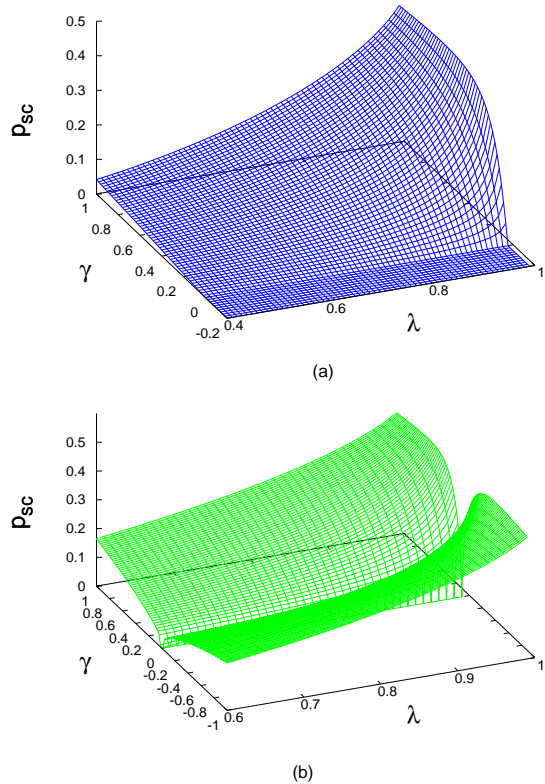}
 \caption{(Color online) (a) Variation of $p_{sc}$ as a function of the system 
         parameters $\lambda$ and $\gamma$ for the BPF channel (b) Variation of $p_{sc}$ 
         as a function of the system parameters $\lambda$ and $\gamma$ for the PF channel.}
 \label{3d}
\end{figure}

The variations of the mutual information $I\left(\rho_{AB}\right)$, classical correlations 
$C\left(\rho_{AB}\right)$ and QD $Q\left(\rho_{AB}\right)$ versus the parametrized time 
$p$ for $(\lambda=0.7,\gamma=\pm0.7)$ have been plotted in Fig. \ref{pfp} in the case of the PF
channel. The dynamics is of type \textit{(ii)} and there is a specific time interval during which the
QD is greater in magnitude than that of the classical correlations. The classical correlations remain 
constant at a value $I\left(\rho_{AB}\right)_{p=1}$, the mutual information of the fully decohered state,
in the parametrized time interval $p_{sc}< p<1$. In the interval $0<p< p_{sc}$, the 
classical correlations decay with time. On the other hand, the quantum correlations, as measured by 
the QD, undergoes a sudden change in its decay rate at $p=p_{sc}$ and goes to zero 
in the asymptotic limit $p\rightarrow1$. One must note that the type of the dynamics in the case of the 
PF channel is unchanged under a change in sign of the anisotropy parameter $\gamma$, the reason of 
which can be easily understood by examining the evolution rules for the coefficients $c_{i}$'s  
from Eq. (\ref{dm}), (\ref{cv}), (\ref{evolve}) and (\ref{kraus}):
 \begin{eqnarray}
 c_{1}(p)&=&c_{1}(0)(1-p)^{2}\nonumber\\
 c_{2}(p)&=&c_{2}(0)(1-p)^{2}\nonumber\\
 c_{3}(p)&=&c_{3}(0) \nonumber\\
 c_{4}(p)&=&c_{4}(0)
\label{pft}
\end{eqnarray}
Both the coefficients $c_{1}$ and $c_{2}$ decay with time identically in the case of the PF channel whereas the 
other coefficients remain independent of time. Since the change in sign of the anisotropy parameter
$\gamma$ only interchanges the values of $c_{1}$ and $c_{2}$, it can not affect the dynamics that the 
classical and quantum correlations undergo in the PF channel. In this case, the sudden change 
in the decay rates of $Q\left(\rho_{AB}\right)$ and $C\left(\rho_{AB}\right)$ at $p=p_{sc}$ is 
understood by noting that in the interval $0<p<p_{sc}$, the QD is determined by 
$Q\left(\rho_{AB}\right)=Q_{2}$ (Eq. (\ref{qd2})) and for $p_{sc}<p<1$, 
$Q\left(\rho_{AB}\right)=Q_{1}$ (Eq. (\ref{qd1})).
In the regime $p_{sc}<p<1$, $C\left(\rho_{AB}\right)$ is given by 
\begin{eqnarray}
 C\left(\rho_{AB}\right)&=&S\left(\rho_{A}\right)+a\log_{2}\frac{a}{a+b}+b\log_{2}\frac{b}{a+b}\nonumber \\
 &&+d\log_{2}\frac{d}{d+b}+b\log_{2}\frac{b}{d+b}
 \label{cc1}   
\end{eqnarray}  
The matrix elements $a$, $b$ and $d$ are functions of only $c_{3}$ and $c_{4}$ which are 
independent of time (Eq. (\ref{pft})). From Eq. (\ref{vne}), $S\left(\rho_{A}\right)$, being a
function of $c_{4}$ only, is also independent of $p$, thereby ensuring the constancy of the 
classical correlations $C\left(\rho_{AB}\right)$. 
The time instant $p_{sc}$ can be determined as a solution of the equation $Q_{1}=Q_{2}$. 
Fig.  \ref{pgl2}(a) shows the variation of $p_{sc}$ as a function of the inverse of the strength
of the transverse field, $\lambda$. The first derivative of $p_{sc}$ w.r.t. $\lambda$ for a 
fixed value of the anisotropy parameter $\gamma$ diverges as the QPT point $\lambda_{c}=1$ is 
approached (although the results for positive $\gamma$ values only have been shown, identical 
result can be obtained for negative $\gamma$ values as well).  Fig. \ref{pgl2}(b) demonstrates 
the variation of $p_{sc}$ versus $\gamma$ for fixed $\lambda$ values. The behaviour of $p_{sc}$
is symmetric across the anisotropy transition point $\gamma=0$. On approaching $\gamma=0$, 
$p_{sc}$ goes to zero. The inset of Fig. \ref{pgl2}(b) shows the variation of the first 
derivative of $p_{sc}$ w.r.t. $\gamma$ at the QCP $\gamma=0$. The landscapes of $p_{sc}$ versus 
the inverse of the transverse-field strength $\lambda$ and the anisotropy parameter $\gamma$
are shown in Fig. \ref{3d}.

\subsection{Further-neighbour quantum correlations at zero temperature}

\begin{figure}
 \includegraphics[scale=0.425]{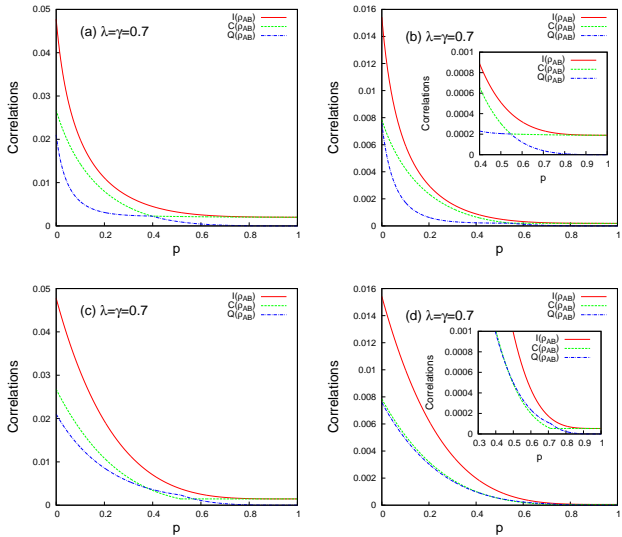}
\caption{(Color online) Decay of the mutual information 
$I\left(\rho_{AB}\right)$ (solid line), the classical correlations $C\left(\rho_{AB}\right)$ (dashed line) and the QD 
$Q\left(\rho_{AB}\right)$ (dot-dashed line) as a function of the parametrized time $p=1-e^{-\theta t}$ for $\lambda=0.7,\;\gamma=0.7$
in the cases of the BPF ($r=2$ (a) and $r=3$ (b)) and the PF ($r=2$ (c) and $r=3$ (d)) channels. Insets of (b) and (d) show 
on a magnified scale the sudden changes in the decay rates of the correlations}
\label{dyn}
\end{figure}

\begin{figure}
 \includegraphics[scale=0.425]{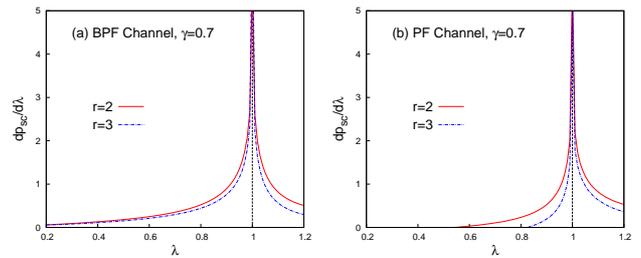}
\caption{(Color online) The divergence of $\frac{dp_{sc}}{d\lambda}$ at the QCP $\lambda_{c}=1$ in the cases of $r=2$ and $r=3$ and 
for the BPF (a) and PF (b) channels with the anisotropy parameter $\gamma=0.7$}
\label{div}
\end{figure}

\begin{figure}
 \includegraphics[scale=0.425]{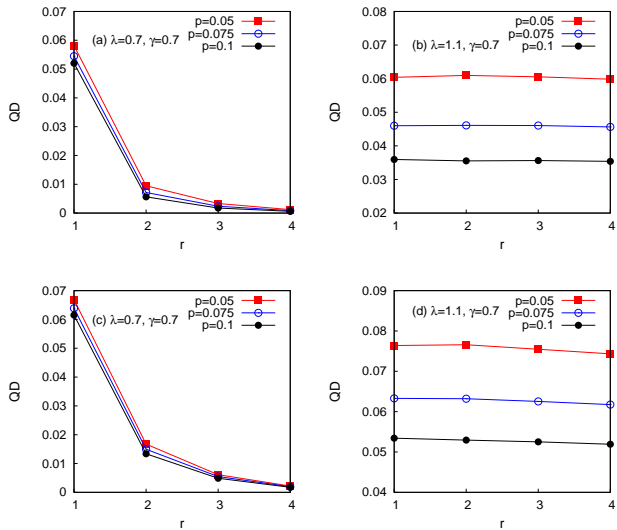}
\caption{(Color online) Decay of the QD as a function of $r$ at three parametrized time points $\left(p=1-e^{-\theta t}\right)$ for the (a),(b) BPF and 
(c),(d) PF channels. For each channel, the decay slows down considerably as the QCP $\lambda_{c}=1$ is crossed. The $p$ values considered are $p=0.05$ (solid squares), 
$p=0.075$ (empty circle) and $p=0.1$ (solid circle). Also $\lambda=0.7$ ((a),(c)), $1.1$ ((b),(d)) and $\gamma=0.7$}
\label{qdecay}
\end{figure}

We have so far considered the spins in a pair to be n.n.s. The dynamics of the QD when the spins are separated by 
further neighbour distances can be investigated following the procedure outlined in Sec. \ref{scheme}. The 
two qubit reduced density matrix $\rho_{sc}(0)$ (Eq. (\ref{evolve})) now involves further-neighbour spin-spin 
correlation functions (Eqs. (\ref{toepliz}) and (\ref{spincor}) with $r>1$). Fig. \ref{dyn} shows the decay of the mutual information 
$I\left(\rho_{AB}\right)$ (solid line), the classical correlations $C\left(\rho_{AB}\right)$ (dashed line) and the QD 
$Q\left(\rho_{AB}\right)$ (dot-dashed line) as a function of the parametrized time $p=1-e^{-\theta t}$ for $\lambda=0.7,\;\gamma=0.7$
in the cases of the BPF ($r=2$ (a) and $r=3$ (b)) and the PF ($r=2$ (c) and $r=3$ (d)) channels with $r=2$ and $3$ corresponding 
respectively to second and third neighbour distances between the spins in a pair.  
Comparing Fig. \ref{dyn} with Figs. \ref{bpfp} and \ref{pfp} one finds that the magnitude of $p_{sc}$
in the case of the PF channel is larger than that in the case of the BPF channel. Our computation 
shows that this feature is obtained when $r=1$ irrespective of whether $\lambda$ is less or greater than 
the QCP $\lambda_{c}=1$. For $r>1$, however, the magnitude of $p_{sc}$ in the case of the BPF channel is 
less than that in the case of the PF channel only for $\lambda<\lambda_{c}$. At $\lambda>\lambda_{c}$, 
the reverse trend is observed. Also, for the same type of channel, BPF or PF, the magnitude of $p_{sc}$ 
shifts towards larger values as $r$, the separation between the two qubits (spins) increases. Fig. \ref{div}
shows the divergence of $\frac{dp_{sc}}{d\lambda}$ at the QCP $\lambda_{c}=1$ for $r=2$ and $3$ and the BPF (a) 
and the PF(b) channels with $\gamma=0.7$. 
Maziero et. al. \cite{mazieroxy,mazieroxy2} have shown that in the case of the 
transverse-field $XY$ chain, the QD
for spin-pairs with $r>1$ is able to signal a QPT whereas pairwise entanglement is non-existent for such distances. They have further 
demonstrated a noticeable change in the decay rate of the QD as a function of $r$ as the system crosses the critical point $\lambda=1$. 
The decay rate is slower for $\lambda>1$, i.e., in the ferromagnetic phase. In the presence of system-environment interactions, 
the QD evolves as a function of time. We select three time points (three different $p$ values) at which we compute the QD for $r=1,2,3$
and $4$.  Figs. \ref{qdecay} (a) and (c) show the results for the BPF and PF channels respectively for $\gamma=0.7$ and $\lambda=0.7$, 
i.e., $\lambda<\lambda_{c}=1$. Figs. \ref{qdecay} (b) and (d) show the corresponding plots when $\lambda$ is $>\lambda_{c}$ with $\lambda=1.1$.
We find that the decay of the QD as a function of $r$ becomes considerably slower as the QCP $\lambda_{c}=1$ is crossed in the case of both the BPF
and PF channels.  

\subsection{Quantum correlations at finite temperatures}

\begin{figure}
\begin{center}
 \includegraphics[scale=0.425]{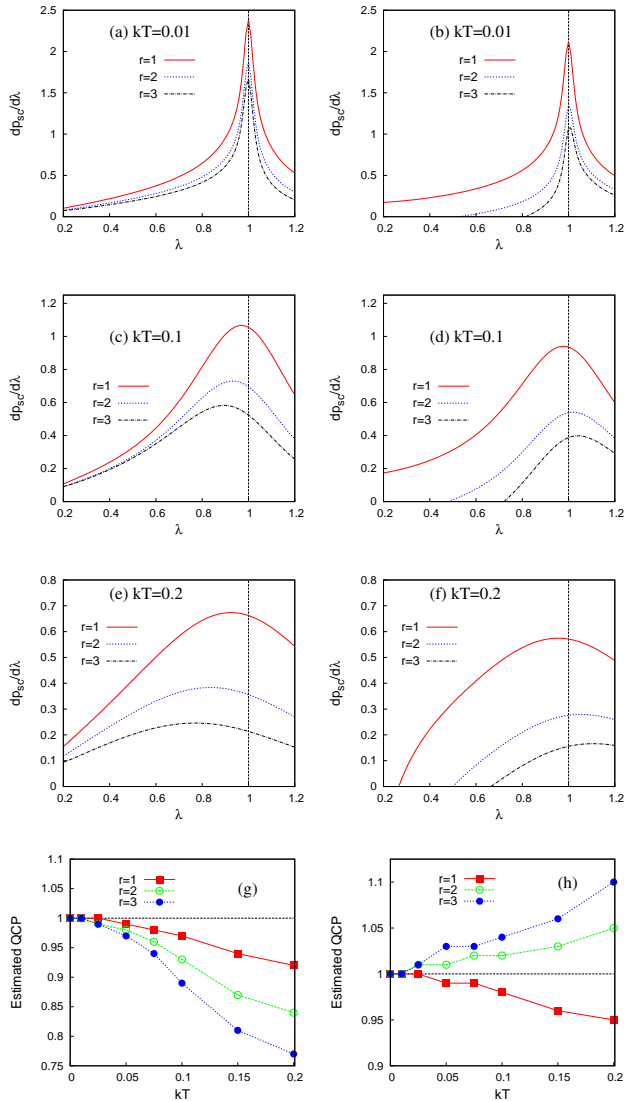}
\end{center}
\caption{(Color online) The variation of $\frac{dp_{sc}}{d\lambda}$ versus $\lambda$ for $r=1$ (solid line), $r=2$ (dotted line) and $r=3$ (dashed line) at temperatures 
$kT=0.01,0.1$ and $0.2$. The first (second) column refers to the BPF (PF) channel. In (g) and (h) the estimated QCPs are plotted against $kT$ for the BPF (g) and PF (h) 
channels respectively and for $r=1$ (solid squares), $r=2$ (empty circles) and $r=3$ (solid circles) with $\gamma=1$.}
\label{fint}
\end{figure}

The dynamics of the QD at finite temperatures ($T\neq0$) and for different values of $r$ are investigated using the procedure described in Sec. \ref{scheme}. 
For infinite chain spin models like the $XXZ$ chain (in the presence and absence of magnetic field) and the transverse-field $XY$ chain, the thermal quantum discord 
(TQD) has been shown to detect the critical points of QPTs at finite temperatures \cite{werlang,werlang2,werlang3}. In fact, the TQD is a better indicator of QPTs than thermodynamic 
quantities and different entanglement measures. Fig. \ref{fint} displays our results when system-environment interactions, which give rise to the dynamics of 
quantum correlations, are taken into account. The first (second) column in the figure pertains to the BPF (PF) channel. The finite temperature calculations 
are carried out for different two-qubit distances, e.g., $r=1,2$ and $3$. An examination of Fig. \ref{fint} shows that at finite temperature, the first derivative 
$\frac{dp_{sc}}{d\lambda}$ no longer diverges at the QCP $\lambda_{c}=1$ but exhibits a maximum which is shifted from the QCP. With increasing $T$ and $r$, the shift is greater 
in magnitude and the maximum becomes more broadened. The last row of Fig. \ref{fint} shows the estimated QCP at $T\neq0$ and for $r=1,2,3$, given by the $\lambda$ value at which 
the maximum of the $\frac{dp_{sc}}{d\lambda}$ versus $\lambda$ plot occurs. In the case of the BPF channel, the estimated QCP is always lower than $\lambda_{c}=1$ for all values of $r$
whereas in the case of the PF channel, the estimated QCP lies above the $\lambda_{c}=1$ value for $r\geq 2$.

\section{Concluding remarks}
\label{conclu}
 The transverse-field $XY$ model in 1d and its derivative model, the transverse-field Ising model, 
are well-known spin models which exhibit QPTs. The models have been extensively studied in recent 
years from the perspective of quantum information theory to quantify the different types of quantum 
correlations like bipartite/multipartite entanglement and quantum discord (QD), present in the ground 
and thermal states of the system, as a function of different model parameters. One significant fact 
which emerges from these studies is that the quantum correlation measures can provide signatures of 
QPTs. Another interesting line of study involves probing the dynamics of different type of 
correlations, both classical and quantum, due to the inevitable interaction of a quantum system 
with its environment resulting in decoherence. In this paper, we study the Markovian time 
evolution of two-qubit states in the Kraus operator formalism due to the local interactions of the 
qubits with independent environments. The two-qubit state is described by the nearest-neighbour 
reduced density matrix of the transverse-field $XY$ model ground state in 1d. The quantum channels, 
representing system-environment interactions, are of three types: bit-flip, bit-phase-flip and 
phase-flip. We investigate the dynamics of the mutual information $I\left(\rho_{AB}\right)$, 
a measure of the total correlations in the system, the classical correlations 
$C\left(\rho_{AB}\right)$ and the QD, $Q\left(\rho_{AB}\right)$ as a function of the parametrized 
time $p$. An earlier study \cite{vedral} identified three different types of dynamics for a 
different set of initial states of which only the type \textit{(ii)} and \textit{(iii)} dynamics 
are observed in the present study. A significant result of our study is to identify quantities, 
$\frac{d p_{sc}}{d\lambda}$ and $\frac{d p_{sc}}{d\gamma}$, 
associated with the type \textit{(ii)} 
dynamics which diverges and becomes discontinuous, respectively, 
as a quantum critical point is approached. The discontinuity occurs as the 
critical point $\gamma=0\;(\lambda\in[0,1])$ of the anisotropy transition is approached and the 
divergence is obtained as the quantum critical point $\lambda=1$ is approached. 
The two different 
types of critical point transitions belong to two different universality classes. We find an 
interesting correspondence between the dynamics of bit-flip and bit-phase-flip channels, namely, 
the dynamics type is interchanged across the $\gamma=0$ line. We provide an explanation for this 
feature and obtain an analytic expression for $p_{sc}$, the parametrized time point at which a 
sudden change in the decay rates of the classical and quantum correlations occur. The 
transverse-field $XY$ model has a phase diagram richer than that of the transverse-field Ising model.
This is reflected in the varied dynamics possible for a single quantum channel in the first case. 
The BPF and BF channels in the case of the transverse-field Ising model are characterized by 
their specific type of dynamics \cite{pal_ising} whereas in the case of the $XY$ model, the dynamics 
type changes across the $\gamma=0$ line. We have further extended our study to finite temperatures and to 
$r$ (the distance of separation between the two qubits described by the reduced density matrix) beyond the 
n.n. distance. In the latter case for $T=0$, one finds that the derivatives $\frac{dp_{sc}}{d\lambda}$ 
diverges at the QCP $\lambda_{c}=1$ as in the case of $r=1$. We have further demonstrated that at a 
specific time point, the decay of the QD as a function of $r$ becomes significantly slower as the QCP 
$\lambda_{c}=1$ is crossed. We have obtained the results for the BPF and the PF channels but the same feature 
holds true for other channels like the amplitude damping channel. We have checked that the decay rate decreases 
continuously as the QCP is approached. At finite temperatures, the TQD is able to provide estimates of critical 
points associated with QPTs, this is true for both n.n. as well as further-neighbour qubits though higher 
temperatures and larger values of $r$ smear out the signatures. The general trends obtained are similar to those derived 
in the case of isolated systems, i.e., when system-environment interactions are ignored. In all our computations, we have not,
however, obtained any evidence of the type \textit{(i)} dynamics mentioned in Sec. \ref{intro}. In the present study, we have considered 
the time evolution of the quantum correlations to be Markovian in nature, an issue of significant interest would be to look for 
signatures of QPTs in the case of non-Markovian time evolution.

\end{document}